\documentclass[aps,prd,preprint,superscriptaddress,tightenlines,nofootinbib]{revtex4}

\usepackage{graphicx}% Include figure files
\usepackage{dcolumn}% Align table columns on decimal point
\usepackage{bm}% bold math

\begin{document}

%\preprint{CLEO CONF 06-10}   % For conference papers
\preprint{CLNS 06/1969}       % for CLNS notes
\preprint{CLEO 06-14}         % for CLNS notes

\title{Branching Fraction for the Doubly-Cabibbo-Suppressed Decay 
$D^{+} \rightarrow K^{+} \pi^{0}$}
 
\author{S.~A.~Dytman}
\author{W.~Love}
\author{V.~Savinov}
\affiliation{University of Pittsburgh, Pittsburgh, Pennsylvania 15260}
\author{O.~Aquines}
\author{Z.~Li}
\author{A.~Lopez}
\author{S.~Mehrabyan}
\author{H.~Mendez}
\author{J.~Ramirez}
\affiliation{University of Puerto Rico, Mayaguez, Puerto Rico 00681}
\author{G.~S.~Huang}
\author{D.~H.~Miller}
\author{V.~Pavlunin}
\author{B.~Sanghi}
\author{I.~P.~J.~Shipsey}
\author{B.~Xin}
\affiliation{Purdue University, West Lafayette, Indiana 47907}
% ...
%\author{(CLEO Collaboration)} %FOR PRD_SPECIAL_CHANGEME
\collaboration{CLEO Collaboration} %FOR PRL,CLNS
\noaffiliation

\date{September 22, 2006}

\begin{abstract} 
We present a measurement of the branching fraction for the 
doubly-Cabibbo-suppressed decay 
$D^{+} \rightarrow K^{+} \pi^{0}$, using 281 $\mathrm{pb}^{-1}$ of data 
accumulated with the CLEO-c detector on the
$\psi$(3770) resonance. We find 
${\cal B} \left( D^{+} \rightarrow K^{+}
\pi^{0}\right) = (2.28 \pm 0.36 \pm 0.15 \pm 0.08) \times
10^{-4}$, 
where the first uncertainty is statistical, the second is systematic, 
and the last error is due to the uncertainty in the reference mode
branching fraction.

\end{abstract}

\maketitle

The Cabibbo-favored hadronic decays of the $c$ quark proceed 
through $c \rightarrow s
W^{+}_{V},\ W^{+}_{V} \rightarrow u \overline d$ 
($W^{+}_{V}$ a virtual $W^{+}$ boson). 
The doubly-Cabibbo-suppressed decays proceed through $c \rightarrow d W^{+}_{V}$,
$W^{+}_{V} \rightarrow u \overline s$, and are expected to be suppressed
by a factor
$|(V_{cd}V_{us})/(V_{cs}V_{ud})|^{2} \approx 2.5 \times 10^{-3}$.
The doubly-Cabibbo-suppressed decay $D^{0} \rightarrow K^{+} \pi^{-}$
was first observed in \mbox{1994 \cite{KPRef}}, and its branching fraction is
now known to good precision ($\pm$2.8\%, relative \cite{PDGValue}). 
Its ratio to the Cabibbo-favored decay 
$D^{0} \rightarrow K^{-} \pi^{+}$ is measured to be 
(3.76 $\pm$ 0.09)$ \times 10^{-3}$ \cite{PDGValue},
in qualitative agreement with the simple expectations.
%%%%%%%Change according to Refree Sep. 14 2006 %%%%%%%%%%%%%%%%%%%%%%%%%%
Very recently BaBar has observed \cite{babar} a second $D \rightarrow
K \pi$ doubly-Cabibbo-suppressed decay $D^{+} \rightarrow K^{+} \pi^{0}$
(charge-conjugate mode $D^{-} \rightarrow K^{-} \pi^{0}$ implied
also, throughout). Here we report confirmation of BaBar's result, with
slightly better accuracy. These measurements can provide insight into
the decay mechanisms for $D \rightarrow K \pi$: the validity of SU(3),
and the roles of the annihilation, exchange, and color-suppressed
spectator diagrams relative to the color-favored spectator diagram
\cite{theoretical1,theoretical2}. A more extensive
picture will be provided by the measurement of the remaining two 
$D \rightarrow K \pi$ doubly-Cabibbo-suppressed decays, 
$D^{+} \rightarrow K^{0} \pi^{+}$ and $D^{0} \rightarrow K^{0} \pi^{0}$.
%%%%%%%%%%%%%%%%%%%%%%%%%%%%%%%%%%%%%%%%%%%%%%%%%%%%%%%%%%%%%%%%%%%%%%%%%%

For this measurement, we have used a 281 $\mathrm{pb}^{-1}$ sample of 
$e^+ e^-$ colliding beam events, collected at a center-of-mass energy
of 3770 MeV. The events were produced
with the \mbox{CESR-c} storage ring and detected with the CLEO-c
detector. The data sample contains about $0.8 \times 10^{6}$ $D^{+}
D^{-}$ events (our target sample), one million $D^{0} \overline D^{0}$ 
events, five million $e^{+} e^{-} \rightarrow u \overline u,\ d \overline d,$ or
$s \overline s$ continuum events, one million $e^{+} e^{-}
\rightarrow \tau^{+} \tau^{-}$ events, and one million $e^{+} e^{-}
\rightarrow \gamma \psi^{\prime}$ radiative return events
(sources of background), as well as Bhabha events, $\mu$-pair
events, and $\gamma \gamma$ events (useful for luminosity
determination and resolution studies).

The CLEO-c detector is a general purpose solenoidal detector which
includes a tracking system for measuring momenta and specific
ionization ($dE/dx$) of charged particles, a Ring Imaging Cherenkov detector
(RICH) to
aid in particle identification, and a CsI calorimeter for detection of
electromagnetic showers. The CLEO-c detector is described in detail
elsewhere \cite{CLEODeter002,CLEODeter003,CLEODeter004}.

The $\psi$(3770) resonance is below the kinematic threshold
for $D \overline D \pi$ production, and so the events of interest, $e^{+}
e^{-} \rightarrow \psi(3770) \rightarrow D \overline D$,
have $D$ mesons with energy equal to the beam energy. 
Having picked the particles being considered to make up
a $D$ meson, following \mbox{Mark III \cite{MARK3}}  we define the two
variables: 

\begin{eqnarray}
\Delta E \equiv \sum_{i} E_{i} - E_{\rm beam},
\end{eqnarray}
and
\begin{eqnarray}
M_{\mathrm {bc}}\equiv \sqrt{E^{2}_{\rm beam} - |\sum_{i} \vec{P}_{i} |^{2}},
\end{eqnarray}
where \noindent $E_{i},\ \vec{P}_{i}$ are the energy and momentum of
each $D$ decay product. For a correct combination of particles,
$\Delta E$ will be consistent with zero, and the beam-constrained mass
$M_{\mathrm {bc}}$ will be consistent with the $D$ mass.

In addition to $D^{+} \rightarrow K^{+} \pi^{0}$, we have studied
the singly-Cabibbo-suppressed decay $D^{+} \rightarrow \pi^{+}
\pi^{0}$, as a higher-rate decay possessing kinematics similar to 
$D^{+} \rightarrow K^{+} \pi^{0}$, and the Cabibbo-favored
decay $D^{+} \rightarrow K^{-} \pi^{+} \pi^{+}$, as a high-rate,
low-background mode used for normalization. We distinguish between
$K^{\pm}$ and $\pi^{\pm}$ using information from the RICH and
$dE/dx$ information from the central drift chamber. We identify
$\pi^{0}$'s via $\pi^{0} \rightarrow \gamma \gamma$, detecting the
photons in the CsI calorimeter. We require that the
calorimeter clusters have a measured energy above 30 MeV, have a
lateral distribution consistent with
that from photons, and are not matched to any charged track. We
require that the $\gamma \gamma$ invariant mass be within 3
standard deviations of the $\pi^{0}$ mass. 
%%%%%%%Change according to Refree Sep. 14 2006 %%%%%%%%%%%%%%%%%%%%%%%%%%
The $\pi^0$ mass resolution is 5.4 MeV (Gaussian width $\sigma$) for
both $D^{+} \rightarrow K^{+} \pi^{0}$ and $D^{+} \rightarrow \pi^{+}
\pi^{0}$. The $\Delta E$ resolution is 14 MeV for $D^{+} \rightarrow
K^{+} \pi^{0}$, 15 MeV for $D^{+} \rightarrow \pi^{+} \pi^{0}$, and
5.6 MeV for $D^{+} \rightarrow K^{-} \pi^{+} \pi^{+}$. The 
$M_{\mathrm{bc}}$ resolution is 1.90 MeV for $D^{+} \rightarrow K^{+}
\pi^{0}$, 1.96 MeV for $D^{+} \rightarrow \pi^{+} \pi^{0}$, and 1.35
MeV for $D^{+} \rightarrow K^{-} \pi^{+} \pi^{+}$.

%%%%%%%%%%%%%%%%%%%%%%%%%%%%%%%%%%%%%%%%%%%%%%%%%%%%%%%%%%%%%%%%%%%%%%%%%

We select candidate combinations that have $\Delta E$ between
$-$40 MeV and +35 MeV for $K^{+} \pi^{0}$ and $\pi^{+} \pi^{0}$,
and between $-$20 MeV and +20 MeV for $K^{-} \pi^{+} \pi^{+}$.
These requirements correspond to roughly 3 standard deviations.
The asymmetric cut for $K^{+} \pi^{0}$ and $\pi^{+}
\pi^{0}$ is due to a low-side tail on $\pi^{0}$ energies, and the
wider window is due to poorer energy resolution. To study background,
we select combinations with $\Delta E$ between $-$100 and $-$50 MeV,
and between +45 and +100 MeV (+50 and +100 MeV for $K^{-} \pi^{+}
\pi^{+}$). When an event contains more than one
$K^+ \pi^0$ combination that passes our $\Delta E$ requirement (a
1.4\% occurrence), we
choose the combination with $\Delta E$ value closest to zero. Multiple
candidates per event for $\pi^+ \pi^0$ and for $K^- \pi^+ \pi^+$ are
comparable in frequency, and are
removed by the same procedure. Thus, we allow only one candidate per
event per decay mode per $D$ charge.
%%%%%%%Change according to Refree Sep. 14 2006 %%%%%%%%%%%%%%%%%%%%%%%%%%
For those multiple candidate events that contain a real $D^{+}
\rightarrow K^{+} \pi^{0}$ decay, Monte Carlo studies indicate that
our algorithm for picking the ``best candidate'' gets the right one 2/3
of the time. Because the algorithm uses $\Delta E$ only, and our
procedure for extracting yield uses a fit to $M_{\mathrm {bc}}$, the
algorithm introduces no bias.
%%%%%%%%%%%%%%%%%%%%%%%%%%%%%%%%%%%%%%%%%%%%%%%%%%%%%%%%%%%%%%%%%%%%%%%%%

%%%%%%%%%%%%%%%%%%%%%%%%%%%%%%%%%%%%%%%%%%%%%%%%%%%%%%%%%%%%%%%%%%%%%%%%%%%%%%%%
\begin{figure}[htbp]

\centerline{
  \includegraphics*[width=6.5in]{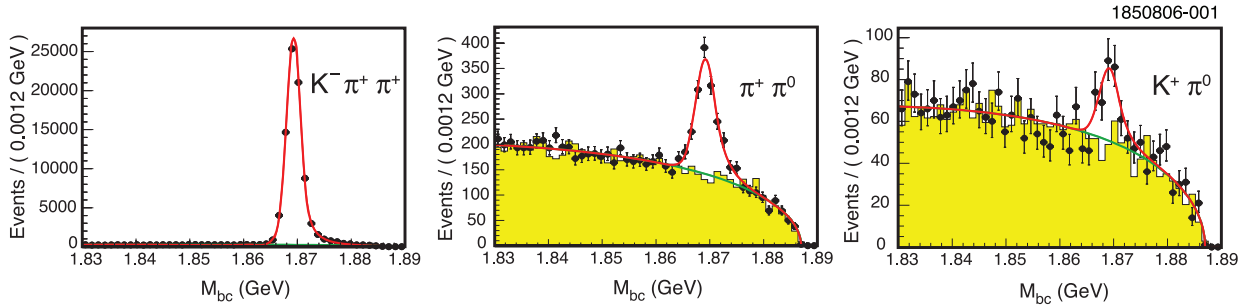} }
\caption{$M_{\mathrm {bc}}$ distributions of $D^+ \rightarrow K^- \pi^+ \pi^+$,  
         $D^+ \rightarrow \pi^+ \pi^0$ and 
         $D^+ \rightarrow K^+ \pi^0$. The points
         are obtained by selecting the $\Delta E$ signal region, the shaded
         histogram is from the $\Delta E$ sidebands, and the lines are
         the fit described in the text. } 
\label{Fig:ST_DATA_Result}
\end{figure}
%%%%%%%%%%%%%%%%%%%%%%%%%%%%%%%%%%%%%%%%%%%%%%%%%%%%%%%%%%%%%%%%%%%%%%%%%%%%%%%%

The $M_{\mathrm {bc}}$ distributions for candidate combinations are shown in
Fig. \ref{Fig:ST_DATA_Result}. The normalization mode $D^{+} \rightarrow K^{-} \pi^{+}
\pi^{+}$ is essentially background-free. The $D^{+} \rightarrow \pi^{+}
\pi^{0}$ mode background is well described by the distribution
obtained from the $\Delta E$ sideband, as is that for the $D^{+}
\rightarrow K^{+} \pi^{0}$ mode. There is a clear peak in $D^{+}
\rightarrow K^{+} \pi^{0}$.

%%%%%%%Change according to Refree Sep. 14 2006 %%%%%%%%%%%%%%%%%%%%%%%%%%
Our Monte Carlo studies indicate that 80\% of the background to $D^{+}
\rightarrow K^{+} \pi^{0}$ comes from continuum events, 11\% from $D
\overline D$ events, 8\% from radiative return events, and 1\%
from $\tau$-pair events. The $\Delta E$
requirement cleanly separates $D^{+} \rightarrow K^{+} \pi^{0}$ and $D^{+}
\rightarrow \pi^{+} \pi^{0}$ decays, so there is no cross-talk between
these modes. There is no evidence for peaking backgrounds.
%%%%%%%%%%%%%%%%%%%%%%%%%%%%%%%%%%%%%%%%%%%%%%%%%%%%%%%%%%%%%%%%%%%%%%%%%

We perform an unbinned maximum likelihood fit to extract signal yields
from the $M_{\mathrm {bc}}$ distributions. 
For the signal, we use a Crystal Ball line shape \cite{CBFunc}, 
which is a Gaussian with a high-side tail. As Monte
Carlo studies show that $D^{+} \rightarrow K^{+} \pi^{0}$ and
$D^{+} \rightarrow \pi^{+} \pi^{0}$ have the same signal shapes, we
have determined the line shape parameters (Gaussian peak location,
Gaussian width, point at which high-side tail begins) from the
$D^{+} \rightarrow \pi^{+} \pi^{0}$  $M_{\mathrm {bc}}$ distribution, and used
them in the fit to the $D^{+} \rightarrow K^{+} \pi^{0}$  $M_{\mathrm {bc}}$
distribution.  We have varied the shape of the high-side tail as
part of the systematic error study.
For the background, we use an
ARGUS function \cite{ArgusFunc}, 
with shape parameter determined from the $\Delta
E$ sideband $M_{\mathrm {bc}}$ distribution, high-end cutoff given by
$E_{\rm{beam}}$, and normalization determined from the fit to the
$\Delta E$ signal region. 
%%%%%%%Change according to Refree Sep. 14 2006 %%%%%%%%%%%%%%%%%%%%%%%%%%
Monte Carlo studies demonstrate that the shape parameter determined
from the $\Delta E$ sideband correctly describes the shape of the
background in the $\Delta E$ signal region. 
We have also performed a fit with the ARGUS shape parameter free
in the fit, and obtained essentially the same result.
%%%%%%%%%%%%%%%%%%%%%%%%%%%%%%%%%%%%%%%%%%%%%%%%%%%%%%%%%%%%%%%%%%%%%%%%%

%%%%%%%%%%%%%%%%%%%%%%%%%%%%%%%%%%%%%%%%%%%%%%%%%%%%%%%%%%%%%%%%%%%%%%%%%%%%%%%%
%    Table list the results from data
\begin{table}[hbtp]
\centering
\caption{The efficiencies (from Monte Carlo, but corrected for 
         $\pi^0$-finding (see below)), 
         fit yields from data, and branching fractions from data. 
         Only statistical uncertainties are included. }
\vspace{0.3cm}
\begin{tabular}{||c||c|c|c||}
\hline \hline
Mode                              &  $\epsilon$ (\%)   &  Signal yield             &  $\mathcal{B}$ (\%)        \\ \hline \hline
$D^+ \rightarrow K^- \pi^+ \pi^+$ &  52.16 $\pm$ 0.16  &  79612 $\pm$ 291   &  9.51 (Input)                 \\ \hline
$D^+ \rightarrow \pi^+ \pi^0$     &  47.65 $\pm$ 0.15  &    964 $\pm$ 54    &  0.1326 $\pm$ 0.0075 \\ \hline
$D^+ \rightarrow K^+ \pi^0$       &  42.30 $\pm$ 0.14  &    148 $\pm$ 23    &  0.0228 $\pm$ 0.0036 \\ 
\hline \hline
\end{tabular}
\label{Tab:ST_DATA_Result}
\end{table}

%%%%%%%%%%%%%%%%%%%%%%%%%%%%%%%%%%%%%%%%%%%%%%%%%%%%%%%%%%%%%%%%%%%%%%%%%%%%%%%%
Results of the fits are shown in Table \ref{Tab:ST_DATA_Result}. 
Also given in Table \ref{Tab:ST_DATA_Result} is
the detection efficiency for each mode, and the branching fractions
obtained for $D^{+} \rightarrow \pi^{+} \pi^{0}$ and $D^{+}
\rightarrow K^{+} \pi^{0}$. Those branching fractions are obtained
by measuring the respective efficiency-corrected yields
relative to that for $D^{+} \rightarrow K^{-} \pi^{+} \pi^{+}$, taking that
branching fraction as $(9.51 \pm 0.34)$\%, which is taken from 
the 2006 Particle Data Group (PDG) value \cite{PDGValue}.
The branching fraction for $D^{+}
\rightarrow \pi^{+} \pi^{0}$ is in good agreement with our
previously-published branching fraction using the same data set,
$(0.125 \pm 0.006 \pm 0.007 \pm 0.004$)\% \cite{BluskPaper}. We
emphasize that these results are {\it not} independent, and 
the value in this paper should {\it not} be used in place of 
the previous result.

We have considered many sources of systematic error to the $D^{+}
\rightarrow K^{+} \pi^{0}$ branching fraction, including: signal
Monte Carlo statistics, track-finding efficiency,
$\pi^{0}$-finding efficiency, particle identification, the $\Delta
E$ requirement, final state radiation, and the uncertainty from
our fitting procedure (background shape, signal shape). The only
ones greater than 1/10 of the statistical error are $\pi^{0}$-finding
efficiency, background shape, and signal shape.

The Monte Carlo simulation of the calorimeter response to photons is
imperfect, particularly in those angular regions where there is
considerable material between the interaction point and the
calorimeter. The Monte Carlo simulation
overestimates the efficiency for detecting $\pi^0$'s. Various
data-Monte Carlo comparisons suggest a correction factor of (0.95
$\pm$ 0.04), which we apply.

The background shape is determined by a fit to the $\Delta E$
sideband data. The error on the shape parameter thus obtained
translates into a $\pm$4.4\% relative error in the $D^{+}
\rightarrow K^{+} \pi^{0}$ branching fraction. The signal shape is
determined by a fit to the $D^{+} \rightarrow \pi^{+} \pi^{0}$
signal.  Uncertainty comes from the determination of Gaussian
width $(\sigma)$, and the point at which non-Gaussian tail sets in
$(\alpha)$. We have obtained the error ellipse in the
determination of these two parameters, and noted the variation in
fitted $D^{+} \rightarrow K^{+} \pi^{0}$ yield as one travels
around this error ellipse. In this way, we obtain a relative systematic
error of $\pm$2.6\%. Note that both the background shape
uncertainty and signal shape uncertainty are really statistical
errors, hence will decrease as additional data are taken.

%%%%%%%Change according to Refree Sep. 14 2006%%%%%%%%%%%%%%%%%%%%%%%%%%%
We have also considered systematic errors to our normalizing mode, 
$D^{+} \rightarrow K^{-} \pi^{+} \pi^{+}$, {\it i.e.}, to the yield and to
the efficiency. Because this mode is essentially background-free,
background shape and signal shape contribute negligible
errors. Kaon particle identification tends to cancel in the ratio to 
$D^{+} \rightarrow K^{+} \pi^{0}$. 
Pion particle identification efficiency is well-modeled by Monte 
Carlo simulation.
Track-finding efficiency -- 3 tracks
in normalizing mode {\it vs.}~1 track in signal mode, with 0.7\% uncertainty
per track -- is the largest error, and is less than 1/10 the overall
statistical error (1.4\% {\it vs.}~16\%).

%%%%%%%%%%%%%%%%%%%%%%%%%%%%%%%%%%%%%%%%%%%%%%%%%%%%%%%%%%%%%%%%%%%%%%%%%

Our final result is

$${\cal B} \left( D^{+} \rightarrow K^{+}
\pi^{0}\right) = (2.28 \pm 0.36 \pm 0.15 \pm 0.08) \times
10^{-4},$$

\noindent where the first error is statistical, the second error
is systematic, and the third error is from the uncertainty in the
$D^{+} \rightarrow K^{-} \pi^{+} \pi^{+}$ branching fraction,
(9.51 $\pm$ 0.34)\% \cite{PDGValue}, used as the normalizing mode.

Our result is in good agreement with the only other measurement of
this branching fraction, BaBar's recent ${\cal B} \left( D^{+}
\rightarrow K^{+} \pi^{0} \right) = (2.52 \pm 0.47 \pm 0.25 \pm 0.08)
\times 10^{-4}$ \cite{babar}. It can be converted to a width, using the PDG
value for the $D^{+}$ lifetime ((1040 $\pm$ 7) $\times 10^{-15}$ s)
\cite{PDGValue}, and compared  with the width for
doubly-Cabibbo-suppressed $D^{0}$ decay $D^{0} \rightarrow K^{+}
\pi^{-}$, using the PDG values for the $D^{0} \rightarrow K^{+}
\pi^{-}$ branching fraction ($(1.43 \pm 0.04)\times10^{-4}$)
\cite{PDGValue} and $D^{0}$ lifetime ($(410.1 \pm 1.5)
\times 10^{-15}$ s)
\cite{PDGValue}.  In this way we obtain

\begin{eqnarray*}
\frac {\Gamma(D^{+} \rightarrow K^{+} \pi^{0})} {\Gamma(D^{0}
\rightarrow K^{+} \pi^{-})} = \frac 
{{\cal B} (D^{+} \rightarrow K^{+} \pi^{0}) \times \tau_{D^0}} 
{{\cal B} (D^{0} \rightarrow K^{+} \pi^{-}) \times \tau_{D^+}}
 =0.63 \pm 0.11 ~.
\end{eqnarray*}

\noindent 
The spectator model diagram, expected to be the dominant contribution, predicts
1/2 for the ratio.  Annihilation and exchange diagrams, which contribute
differently to the two decays, can shift the ratio away from 1/2.  Our result,
and the BaBar result \cite{babar}, suggest that such a shift is small.

    In summary, we have measured the branching fraction for $D^+ \rightarrow K^+ \pi^0$ to
be $(2.28 \pm 0.36 \pm 0.15 \pm 0.08) \times
10^{-4}$, in agreement with the only other
measurement of that branching fraction, and of comparable accuracy.

We gratefully acknowledge the effort of the CESR staff 
in providing us with excellent luminosity and running conditions. 
D.~Cronin-Hennessy and A.~Ryd thank the A.P.~Sloan Foundation. 
This work was supported by the National Science Foundation,
the U.S. Department of Energy, and 
the Natural Sciences and Engineering Research Council of Canada.


\begin{thebibliography}{99}
%%%%%%%%%%%%%%%%%%%%%%%%%%%%%%%%%%%%%
%\bibitem{CLEODeter001} G. Bonvicini $et ~ al$. (CLEO Collaboration), 
%                       Phys. Rev. D \textbf{70}, 112004 2004 [hep-ex/0411050].
\bibitem{KPRef}        D. Cinabro $et ~ al$. (CLEO Collaboration),
                       Phys. Rev. Lett. \textbf{72}, 1406 (1994). 
\bibitem{PDGValue}     W.-M. Yao $et ~ al$. (Particle Data Group),
                       J. Phys. G \textbf{33}, 1 (2006).
\bibitem{babar}        B. Aubert $et ~ al$. (BaBar Collaboration),
                       Phys. Rev. D {\bf 74}, 011107(R) (2006).
\bibitem{theoretical1} L.-L. Chau and H.-Y. Cheng, Phys. Lett. B
                       \textbf{333}, 514 (1994)
\bibitem{theoretical2} C.-W. Chiang and J.L. Rosner, Phys. Rev. D
                       \textbf{65}, 054007 (2002). 
\bibitem{CLEODeter002} Y. Kubota $et ~ al$. (CLEO Collaboration), Nucl. Instrum. Methods
                       Phys. Res., Sect. A \textbf{320}, 66 (1992).
\bibitem{CLEODeter003} D. Peterson $et ~ al$., Nucl. Instrum. Methods
                       Phys. Res., Sect. A \textbf{478}, 142 (2002).
\bibitem{CLEODeter004} M. Artuso $et ~ al$., Nucl. Instrum. Methods
                       Phys. Res., Sect. A \textbf{554}, 147 (2005).
\bibitem{MARK3}        J. Adler $et ~ al$. (Mark III Collaboration),
                       Phys. Rev. Lett. \textbf{62}, 1821 (1989).
\bibitem{CBFunc}       T. Skwarnicki, Ph.D thesis, Institute for
                       Nuclear Physics, Krakow, Poland, 1986.
\bibitem{ArgusFunc}    H. Albrecht $et ~ al$. (ARGUS Collaboration), Phys. Lett. B \textbf{229},
                       304 (1989).
\bibitem{BluskPaper}   P. Rubin $et ~ al$. (CLEO Collaboration),
                       Phys. Rev. Lett. \textbf{96}, 081802 (2006).
%%%%%%%%%%%%%%%%%%%%%%%%%%%%%%%%%%%%%
\end{thebibliography}
\end{document}